\documentclass[structabstract]{aa}  

\usepackage{graphicx}
\usepackage{txfonts}
\usepackage{lscape}
\usepackage{rotating}

\newcommand{\vel}{{\mathrm v}}

\begin{document}

   \title{Multi-wavelength diagnostics of accretion in an X-ray selected sample of CTTSs}
   \titlerunning{Multi-wavelength diagnostics of accretion in CTTSs}

   \author{R. L. Curran
          \inst{1}\fnmsep\inst{2}
          \and
          C. Argiroffi\inst{3}\fnmsep\inst{2}
          \and
          G. G. Sacco\inst{4}\fnmsep\inst{2}
          \and
          S. Orlando\inst{2}
          \and
          G. Peres\inst{3}\fnmsep\inst{2}
          \and
          F. Reale\inst{3}\fnmsep\inst{2}
          \and
          A. Maggio\inst{2}
          }

   \institute{Present Address: Department of Physics, Rochester Institute of Technology, 84 Lomb Memorial Drive, Rochester, NY, 14623, USA\\
              \email{rlcsps@rit.edu}
         \and INAF - Osservatorio Astronomico di Palermo, Piazza del Parlamento 1,
              90134 Palermo, ITALY
\and
Dipartimento di Scienze Fisiche ed Astronomiche - Universit\`{a} degli Studi di Palermo, Piazza del Parlamento 1, 90134 Palermo, ITALY
\and
Chester F. Carlson Center for Imaging Science, Rochester Institute of Technology, 54 Lomb Memorial Drive, Rochester, NY 14623, USA\\}
   \authorrunning{Curran et al}
   \date{Received ; accepted }

 
  \abstract
  {High resolution X-ray spectroscopy has revealed soft X-rays from
    high density plasma in Classical T-Tauri stars (CTTSs), probably
    arising from the accretion shock region.  However, the mass
    accretion rates derived from the X-ray observations are
    consistently lower than those derived from UV/optical/NIR
    studies.}
{We aim to test the hypothesis that the high density soft X-ray
  emission is from accretion by analysing optical accretion tracers from
  an X-ray selected sample of CTTSs in a homogeneous manner.}
{We analyse optical spectra of the X-ray selected sample of CTTSs and
  calculate the accretion rates based on measuring H$\alpha$, H$\beta$,
  H$\gamma$, \ion{He}{ii} 4686\AA\,, \ion{He}{i} 5016\AA\,, \ion{He}{i}
  5876\AA\,, \ion{O}{i} 6300\AA\ and \ion{He}{i} 6678\AA\ equivalent
  widths. In addition, we also calculate the accretion rates based on
  the full width at 10\% maximum of the H$\alpha$ line. The different
  optical tracers of accretion are compared and discussed. The derived
  accretion rates are then compared to the accretion rates derived from
  the X-ray spectroscopy.}
{We find that, for each CTTS in our sample, the different optical
  tracers predict mass accretion rates that agree within the errors,
  albeit with a spread of $\approx 1$ order of magnitude.  Typically,
  mass accretion rates derived from H$\alpha$ and \ion{He}{i} 5876\AA\
  are larger than those derived from H$\beta$, H$\gamma$ and
  \ion{O}{i}. In addition, the H$\alpha$ full width at 10\%, whilst a
  good \em indicator \em of accretion, may not accurately \em measure
  \em the mass accretion rate. When comparisons of the optical mass
  accretion rates are made to the X-ray derived mass accretion rates,
  we find that: a) the latter are always lower (but by varying
  amounts); b) the latter range within a factor of $\approx 2$ around
  $2 \times 10^{-10}$ M$_{\odot}$\,yr$^{-1}$, despite the fact that
  the former span a range of $\approx 3$ orders of magnitude. We
  suggest that the systematic underestimation of the X-ray derived
  mass accretion rates could depend on the density distribution inside
  the accretion streams, where the densest part of the stream is not
  visible in the X-ray band because of the absorption by the stellar
  atmosphere. We also suggest that a non-negligible optical depth of
  X-ray emission lines produced by post-shock accreting plasma may
  explain the almost constant mass accretion rates derived in X-rays
  if the effect is larger in stars with larger optical mass accretion
  rates.  }
{}

   \keywords{accretion, accretion disks --
             circumstellar matter --
             stars: pre-main sequence --
             techniques: spectroscopic
               }

\maketitle

\section{Introduction}

The process of stellar mass accretion is an important aspect of star
formation that is still to be fully understood. Not only is accretion
responsible for building up the young star to its final mass, but it
is also responsible for powering the mass outflows observed from such
systems, which in turn remove the excess angular momentum, and prevent
the star from spin-up. In addition, understanding the mass accretion
rate will significantly impact on the understanding of the inner disk,
disk evolution and the eventual formation of planets. Stellar
accretion can be briefly summarised as follows: material passes from
the envelope through the accretion disk -- which is truncated at a
radius $R_{\mathrm{in}}$ due to the strong magnetic field of the central
star. At this inner region of the disk, the material then flows along
the star-disk magnetic field lines (flux tubes) -- at $\sim$free-fall
velocity -- onto the central star, where a strong shock is formed as the
accreting material impacts the stellar surface.

UV veiling, H$\alpha$, Pa$\beta$, Br$\gamma$, \ion{O}{i}, \ion{Ca}{ii}
and \ion{He}{i} emission lines have all previously been used to probe
the accretion scenario in young stars and Brown Dwarfs (e.g. Natta et
al. \cite{natta04,natta06}; Mohanty et al. \cite{subu03,subu05}). Such
studies are based on the hypothesis that the UV veiling arises from
the hot shock region (e.g. Calvet \& Gullbring \cite{calvet}), whereas
the broad permitted emission lines exhibited by CTTSs arise from the
infalling magnetospheric flow (Muzerolle et
al. \cite{muzerolle98}). The \ion{Ca}{ii} lines have been found to be
particularly good infall tracers (Muzerolle et al. \cite{muzerolle98};
Mohanty et al. \cite{subu05}), with the flux strongly correlated to
the accretion rate, although other lines such as H$\alpha$ and
\ion{He}{i} 5876\AA\, also show good correlations to the accretion
rate (Herczeg \& Hillenbrand \cite{herczeg}).

For the last few decades it has been known that pre-main
sequence stars have strong X-ray emission (Feigelson \& DeCampli
\cite{feigelson81}). This X-ray emission was thought to have
the same origin as that of main sequence stars: low-density plasma
($n_{\mathrm e} \sim 10^{10}$ cm$^{-3}$ enclosed in coronal loop
structures and heated to temperatures of $T \sim 10^{6}-10^{7}$
K (Feigelson \& Montmerle \cite{feigelsonmont99}). More recently,
high resolution X-ray spectroscopy has revealed the presence of soft
($E<0.7$ keV) X-ray emission, originating from high density ($n>10^{11}$
cm$^{-3}$) plasma at temperatures of $\sim$3 MK (see Telleschi et
al. \cite{telleschi07}, and references therein). It has been proposed that
this soft X-ray emission is due to mass accretion. This interpretation
is based on a simple model: assuming the accretion flow has free-fall
velocity $\vel \sim 500$ km\,s$^{-1}$, it becomes heated up by the shock
(due to the impact of the accreting material with the stellar surface)
at a temperature of $T \sim (3/16)(\mu m_{\rm H} \vel^{2})\sim 3 \times
10^6$ K, and then cools down radiatively (Gullbring \cite{gullbring94};
Calvet \& Gullbring \cite{calvet}; Lamzin \cite{lamzin}; Gunther
et al. \cite{gunther07}; Brickhouse et al. \cite{brickhouse}) producing
strong X-ray emission.

Supporting this interpretation is the fact that this plasma component
i) has never been observed in non-accreting stars, and ii) is too
dense to have a coronal origin. This interpretation is also supported
by time-dependent models of radiative accretion shocks in CTTSs
(Koldoba et al. \cite{Koldoba2008MNRAS}; Sacco et al. \cite{germano};
\cite{sacco10}; Orlando et al. \cite{Orlando2010A&A}). In particular,
Sacco et al. (\cite{germano}) carried out a detailed hydrodynamic
modeling of the interaction between the accretion flow and the stellar
chromosphere, synthesizing the high resolution X-ray spectrum, as it
would be observed with the Reflection Grating Spectrometers (RGS) on
board the XMM-Newton satellite. They found an excellent agreement
between predicted and observed X-ray spectra, supporting once again
the idea that this X-ray emission originates mainly from accretion
shocks.

However, there are observational results that are difficult to reconcile
with this framework. In particular, the mass accretion rates derived
from X-rays are usually underestimated if compared to accretion rates
derived from optical and UV data. As an example, for TW Hya, X-rays
indicate $\dot{M} \sim 1 \times 10^{-11} M_{\odot}$\,yr$^{-1}$ (Stelzer
et al. \cite{beate}), while H$\alpha$ and UV provide $5 \times 10^{-10}$
M$_{\odot}$\,yr$^{-1}$ (Muzerolle et al. \cite{muzerolle00}). The
same mismatch holds for BP Tau: $\dot{M} \sim 9 \times 10^{-10}$
M$_{\odot}$\,yr$^{-1}$ from X-rays (Schmitt et al. \cite{schmitt}),
and $3 \times 10^{-8}$ M$_{\odot}$\,yr$^{-1}$ from UV (Gullbring et
al. \cite{gullbring}). To date, only 3 such comparisons have been
made in the literature (TW Hya; BP Tau; MP Mus, Argiroffi et al
\cite{costanza09}), moreover, each of these comparisons has used
differing UV/optical accretion measures and differing means of calculating
the X-ray accretion rate. In order to confirm that the soft X-ray emission
arises from the accretion process, and understand the discrepancies
between the optical and X-ray accretion rates, we have homogeneously
analysed optical data for all the CTTSs for which high-resolution X-ray
spectra, with good S/N in the [0.5, 1.0] keV range, have been
gathered. We compare the X-ray derived mass accretion rates to those
measured from the H$\beta$, \ion{He}{i} 5876\AA, \ion{O}{i} 6300\AA,
H$\alpha$ equivalent width and H$\alpha$ full-width at 10\%.

The structure of the paper is as follows. In section 2 we present our
sample of CTTSs previously observed with Chandra and XMM. We briefly
discuss the parameters of each CTTS (summarised in Table \ref{param}),
along with the previously published X-ray data. In section 3 we
discuss the data reduction and analysis of the optical data
(subsection 3.1), and the derivation of mass accretion rates from the
X-ray data (subsection 3.2). In section 4.1 we draw comparisons
between the different optical tracers of accretion and discuss the
mass accretion rates derived from these tracers. In section 4.2 we
analyse and discuss the variability of TW Hya. In section 4.3 we
compare the optical and X-ray mass accretion rates and discuss
possible scenarios to explain our results, and in section 5 we present
our conclusions.

\begin{sidewaystable*}
\vspace{15cm}
\caption{List of sources and their adopted parameters.}
\label{param}
\centering
\begin{tabular}{lcccccccccccl}
\hline \hline
Name               & RA         &  Dec      & Dist. & Mass         & Radius     & Age & Sp Type & Mag. & A$_{\mathrm{v}}$    & Incl.   & Binarity\tablefootmark{a} & References \\
~                  &(J2000)     & (J2000)   & (pc)     & (M$_{\odot}$) & (R$_{\odot}$)& Myr& ~             & (Johnson V)   & (mag)     &  (\degr)  & ~ & ~          \\
\hline
Hen 3-600 & 11 10 28.1 & -37 31 51 & 45       & 0.2      & 0.9     &8     & M4Ve/M4Ve      & 12.04         & 0.7   & $\sim0$   &  BS  & 1, 2, 3, 4, 5, 6 \\
TW Hya    & 11 01 51.9 & -34 42 17 & 56       & 0.7      & 1.0     &8     & K7        & 11.27         & 0     & 7         &  SS  & 1, 6, 7, 8, 9, 10, 11, 12 \\
RU Lup    & 15 56 42.3 & -37 49 15 & 140      & 0.8      & 1.7     &0--3  & K7        & 11.55         & 0.1   & 24        &  SS  & 7, 9, 13, 14, 15, 16 \\
BP Tau    & 04 19 15.8 & +29 06 27 & 140      & 0.8      & 2.0     &1.9   & K7        & 12.13         & 0.5   & 45        &  SS  & 9, 17, 18, 19, 20, 21, 22 \\
V4046 Sgr & 18 14 10.5 & -32 47 34 & 72       & 0.86     & 1.16    &12    & K5Ve/K7Ve & 10.69         & 0     & 35        &  BS  & 9, 23, 24, 25 \\
MP Mus    & 13 22 07.5 & -69 38 12 & 86       & 1.2      & 1.3     &6--7  & K1IVe     & 10.44         & 0.17  & 32        &  SS  & 9, 26, 27, 28 \\
V2129 Oph & 16 27 40.3 & -24 22 03 & 120      & 1.35     & 2.4     &2     & K5        & 12.28         & 0.3   & 45        &  SS  & 3, 9, 29, 30 \\
T Tau N   & 04 21 59.4 & +19 32 06 & 140      & 2.4      & 3.6     &2.7   & K0        & 9.88          & 1     & 13        &  TS  & 8, 15, 16, 20, 31\\
\hline
\end{tabular}

\tablefoottext{a}{ SS = single star; BS = binary system; TS = triple system.}
\tablebib{
(1) Muzerolle et al. \cite{muzerolle00}; 
(2) Torres et al. \cite{torres00}; 
(3) Geoffray \& Monin \cite{geoffray}; 
(4) Huenemoerder et al. \cite{huenemoerder}; 
(5) Kastner et al. \cite{kastner97}; 
(6) Song et al. \cite{song};
(6) Herczeg \& Hillenbrand \cite{herczeg}; 
(7) Qi et al. \cite{qi04}; 
(8) Lasker et al. \cite{lasker}; 
(9) Wichman et al. \cite{wichman}; 
(10) Raassen \cite{raasan}; 
(11) Brickhouse et al. \cite{brickhouse}; 
(12) Robrade \& Schmitt \cite{robrade}; 
(13) Stempels, Gahm \& Petrov \cite{stempelsgahmpetrov}; 
(14) Stempels \& Piskunov \cite{stempels};
(15) Comer\'{o}n et al. \cite{comeron}; 
(15) G\"udel et al. \cite{gudel}; 
(16) Grankin et al. \cite{grankin}; 
(17) Costa et al. \cite{costa}; 
(18) Donati et al. \cite{donati08};  
(19) Kenyon et al. \cite{kenyon}; 
(20) Kenyon \& Hartmann \cite{kenyonhart}; 
(21) Stempels \& Gahm \cite{stempelsgahm}; 
(22) Hutchinson et al. \cite{hutchinson}; 
(23) Kastner et al. \cite{kastner08}; 
(26) Mamajek et al. \cite{mamajek}; 
(27) Cortes et al. \cite{cortes09};
(28) Torres et al. \cite{torres08}; 
(29) Donati et al. \cite{donati}; 
(30) Wilking et al. \cite{wilking}; 
(31) Bertout et al. \cite{bertout}.           
}

\end{sidewaystable*}

\section{The Sample}

The sample consists of all the CTTSs currently observed with
high-resolution X-ray spectroscopy (either with Chandra, or RGS on
XMM-Newton) and for which \ion{O}{vii} triplet lines have been
measured, and which also have high resolution optical echelle
spectroscopy available in the data archives. The X-ray data have
previously been published, with the published fluxes of the
\ion{O}{vii} used to derive the mass accretion rate for each star. The
parameters for each source, used in our calculations are given in
Table~\ref{param}. Below are brief descriptions of each source.

{\em \object{Hen 3-600}} is a multiple star system and a member of the
TW Hydra Association, at a distance of 45 pc. This is an average
  distance taken from Huenemoerder et al. (\cite{huenemoerder}), who
  used the photometric distance from Kastner et al. (\cite{kastner97})
  and the typical distance adopted for stars belonging to the TW Hya
  association. The primary components (A and B) of this system are
separated by 1\farcs4, and component A is surrounded by a dusty disc
(Jayawardhana et al. \cite{rayjay99}). Component A is also inferred
(by Huenemoerder et al. \cite{huenemoerder}) to be almost pole-on,
from a large mid-IR excess and a negligible optical reddening (based
on a B$-$V = 1.52 which is nearly that of an unreddened M3
photosphere; Johnson \cite{johnson66}). Huenemoerder et
al. (\cite{huenemoerder}) discuss the X-ray Chandra observations of
Hen 3-600, and find high density plasma, and a larger `soft excess'
for component A, than B.

{\em \object{TW Hya}} is one of the closest known CTTSs, at a distance
of only $\sim$ 56 pc (Wichmann et al. \cite{wichman}). It has a
mass of 0.7 M$_{\odot}$, a radius of 1 R$_{\odot}$ (Muzerolle et
al. \cite{muzerolle00}) and is orientated so that it is seen nearly
pole-on (Kastner et al. \cite{kastner97}). Previous X-ray data have
been published by Kastner et al. (\cite{kastner02}), Raassen
  (\cite{raasan}), Brickhouse et al. (\cite{brickhouse}) and Stelzer
\& Schmitt (\cite{beate}), who found a strong soft X-ray emission
produced by high density plasma.

{\em \object{RU Lup}} is located at a distance of 140 pc (Hughes et
al. \cite{hughes93}). It has a mass and radius of 0.8 M$_{\odot}$ and
1.7 R$_{\odot}$ respectively (Stempels \& Piskunov
  \cite{stempels}), it suffers little absorption (Herczeg et
al. \cite{herczeg05}) and is suggested to be viewed almost pole-on
(Stempels \& Piskunov \cite{stempels}). XMM data are discussed by
Robrade \& Schmitt (\cite{robrade}), who find cool, high density
  plasma which they conclude indicates an accretion shock origin.

\begin{table*}
\caption{Optical archival data information for our sample of CTTSs.}
\label{obs}
\centering
\begin{tabular}{lccccc}
\hline \hline
Name               & Telescope & Instrument & Observation Date & Exposure time \\
~                  & ~         & ~          & (yyyymmdd)       & (s)           \\
\hline
Hen 3-600 & ESO 2.2m  & FEROS      & 20040512         & 900           \\
TW Hya    & ESO 2.2m  & FEROS      & 20070426         & 900           \\
RU Lup    & VLT       & UVES       & 20050814         & 900           \\
BP Tau    & TNG       & SARG       & 20071220         & 3600          \\
V4046 Sgr & VLT       & UVES       & 20050815         & 900           \\
MP Mus    & ESO 2.2m  & FEROS      & 20060418         & 3000          \\
V2129 Oph & VLT       & UVES       & 20020417         & 900           \\
T Tau     & TNG       & SARG       & 20061130         & 3600          \\
\hline
\end{tabular}
\end{table*}

{\em \object{BP Tau}} is a CTTS in the Taurus-Auriga molecular cloud at a
distance of 140 pc (Kenyon et al. \cite{kenyon}). It has a spectral
type of K7, and a mass of 0.8 M$_{\odot}$ (Kenyon \& Hartmann
\cite{kenyonhart}; G\"{u}del et al. \cite{gudel}). The XMM-Newton
spectra were previously discussed by Schmitt et al. (\cite{schmitt})
and Robrade \& Schmitt (\cite{robrade06}). They report the presence
of high density soft X-ray emission from the \ion{O}{vii} triplet,
suggesting this could be due to accretion shocks.

{\em \object{V4046 Sgr}} is a nearby spectroscopic binary CTTS that is
isolated from any dark cloud or molecular cloud, it has negligible
extinction, and there is evidence for a circumstellar disk (Hutchinson et
al. \cite{hutchinson}). The period is well determined to be 2.4213459 days
(Stempels \& Gahm \cite{stempelsgahm}) and the separation is estimated
to be $\sim$ 10 R$_{\odot}$ (Quast et al. \cite{quast}). The observed
spectral energy IR distribution is consistent with the disk having
an inner radius of 1.8AU, therefore the disk is circumbinary. The
inclination of the system is $\sim$35--45\degr. Chandra data of V4046
Sgr is presented by G\"{u}nther et al. (\cite{gunther}), who found the
X-ray emission is due to high density plasma.

{\em \object{MP Mus}} is a K1 IVe type star located in the Lower
Centaurus Crux (LCC) association at a distance of $\sim$ 86 pc. It is
known to have a dusty disk (Mamajek et al. \cite{mamajek}; Silverstone
et al. \cite{silverstone}) with a dust mass of $\sim 5 \times
10^{-5}$ M$_{\odot}$ (Carpenter et al. \cite{carpenter}). Batalha et
al. (\cite{batalha}) studied the variability in B, V, R and I bands,
and found that it has a low variability for a CTTS. XMM data for MP Mus
have previously been published by Argiroffi et al. (\cite{costanza}),
who found evidence for a high density plasma responsible for the soft
X-ray emission.

{\em \object{V2129 Oph}} is the brightest CTTS in the $\rho$ Oph cloud,
at a distance of 120 pc (Lombardi, Lada \& Alves (\cite{lombardi})). It
is a distant binary, with its very low-mass companion (possibly a
brown dwarf) being about 50 times fainter than the CTTS in the V band
(Geoffray \& Monin \cite{geoffray}). The CTTS is inferred to have a mass
of 1.35 M$_{\odot}$ and a radius of 2.4 R$_{\odot}$. Chandra data
have recently been obtained (Argiroffi et al. in prep) but not published
yet. We include the optical analysis in this paper for easy comparison
once the X-ray data is published.

{\em \object{T Tau} N} is another CTTS located in Taurus-Auriga. It is
 optically visible part of a triple system, along with T Tau Sa
  and Sb which are the ``infrared companions''. It has a spectral
type of K0 (Kenyon \& Hartmann \cite{kenyonhart}) and a mass of
2.7 M$_{\odot}$ (G\"{u}del et al. \cite{gudel}). It is oriented such
that it is seen almost pole-on (Solf \& B\"{o}hm \cite{solfandbohm};
Akeson et al. \cite{akeson}). X-ray data from XMM and Chandra have
been discussed in G\"{u}del et al. (\cite{gudel}) who found a `soft
excess' but no evidence of high density plasma.

\section{Data Analysis}

\subsection{Optical Data Analysis}

We retrieved high resolution echelle spectra spanning the optical
wavelength range from the archives. No single instrument had observed
the entire sample (in part, due to their locations). Data from three
instruments, FEROS, on the 2.2 m ESO telescope in Chile, UVES, on the
VLT in Chile and SARG on the TNG in La Palma were used in this
analysis. Details of the telescope, instrument, observation date and
exposure times for each source can be found in
Table~\ref{obs}. Relevant calibration files were downloaded with each
dataset and used in the data reduction process.

\begin{table*}
\caption{Measured equivalent widths for the accretion tracers\tablefootmark{a}, along
  with the measured full width at 10\% for the H$\alpha$
  line.}
\label{tab:linewidths}
\centering
\begin{tabular}{lccccccccc}
\hline \hline
\\
Source	& H$\gamma$ 4340\AA & \ion{He}{ii} 4686\AA & H$\beta$ 4861\AA	& \ion{He}{i} 5016\AA	& \ion{He}{i} 5876\AA	& \ion{O}{i} 6300\AA	& H$\alpha$ 6563\AA 	& H$\alpha$ FW10\%	& \ion{He}{i} 6678\AA    \\
	& \AA	            & \AA	           & \AA	        & \AA	                & \AA	                & \AA	                & \AA	                & km\,s$^{-1}$  	& \AA            \\
\hline
Hen 3-600&	-7.51	    &  ...	           & -7.93	        & -0.32	                & -0.96	                & -0.37	                & -21.81	        & 282.29	        & -0.37          \\
TW Hya	& -18.49	    & -0.48	           & -31.9	        & -0.32	                & -3.16	                & -0.54	                & -141.58	        & 407.55	        & -0.7           \\
RU Lup	& NA	            &  NA	           & -28.34	        & -9.25	                & -4.55	                & -1.17	                & -68.95	        & 578.42	        & -1.88          \\ 
BP Tau	& NA	            & -0.6	           & -19.82	        & -0.27	                & -1.08	                & -0.37	                & -58.63	        & 413.41	        & -0.31          \\
V4046 Sgr& NA	            &  NA	           & -5.97	        &  ...	                & -0.41	                & -0.12	                & -32.66	        & 515.19	        & ...            \\
MP Mus	& -1.34	            &  ...	           & -3.16	        &  ...	                & -0.3	                & -0.23	                & -23.47	        & 502.9	                & 0.16           \\
V2129 Oph& NA	            &  NA	           & -2.54	        &  ...                  & -0.33	                & -0.13	                &-12.07	                & 274.24	        & -0.04          \\
T Tau	& NA               & -0.1	           & -12.05	        & -0.6	                & -0.55	                & -0.7	                & -42.45	        & 419.71	        & 0.08           \\
\hline
\end{tabular}
\tablefoottext{a}{ Negative values represent emission. Note some sources show
  lines in absorption. There are no measurements where the
  spectra show no indication of a line present. NA indicates where the wavelength
  of the line was not covered in the observation.}
\end{table*}

The data from the three instruments were reduced in the standard
manner using IRAF routines to de-bias, flatfield and extract the
spectra. Wavelength calibrations were carried out based on ThArNe arcs
for FEROS, ThAr arcs for UVES and Thorium arcs for SARG. 

Line luminosities, $L_{\mathrm{line}}$, have been calculated from the
equivalent width of each emission line (see
table~\ref{tab:linewidths}). The continuum fluxes have been estimated
from the V magnitudes of the stars, following a method similar to that
used by Mohanty et al. (\cite{subu05}) and Dahm (\cite{dahm}).
Specifically, the continuum flux at 5500\,\AA\, was calculated using
the extinction corrected observed V magnitude, the distance, and the
photometric zero point of the Johnson V filter (at the central
wavelength 5500\,\AA) of 3.75$\times$ 10$^{-9}$
\,erg\,cm$^{-2}$\,s$^{-1}$\,\AA$^{-1}$ (Mitchell \& Johnson
\cite{mj}). The continuum flux at the wavelength of each of the
analysed emission lines also required correction for the stellar
spectral response -- which was measured using the Pickles
Spectrophotometric Atlas of Standard Stellar Spectra (Pickles
\cite{pickles}) in conjunction with the observed stellar spectral
type. It is worth noting that using the observed V magnitude we take
into account the continuum excess due to the accretion process.

The mass accretion rates were calculated following the empirical
relation between accretion and line luminosities found and described
by Herczeg \& Hillenbrand (\cite{herczeg}):

 \begin{figure*}
   \centering
   \includegraphics[width=15cm]{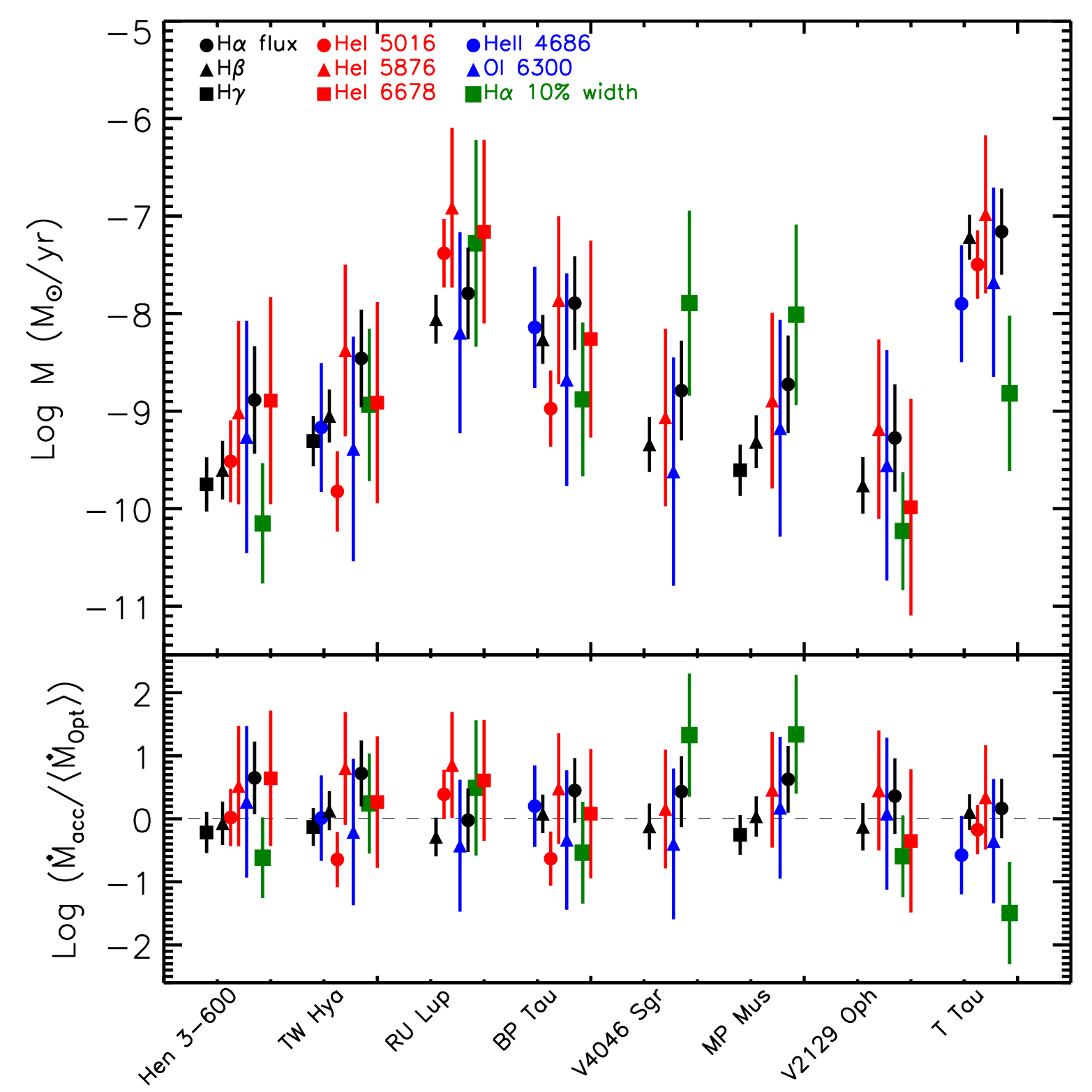}
   \caption{{\em Top Panel:\em} Plot of the different mass accretion
     rates from different accretion tracer emission lines, for each of
     the stars in the sample. The stars are in order of increasing
     mass, showing no relation between mass and mass accretion
     rate. {\em Bottom Panel:\em} Plot showing the ratio of each mass
     accretion rate estimate to the optical mean of the accretion rate
     for each star. The different symbols/colors represent different
     accretion rate tracers (see upper left corner in the top panel).}
   \label{opticalcomp}
   \end{figure*}

\begin{equation}
\label{line}
\log L_{\rm acc} = a + b \log L_{\rm line}
\end{equation}

\noindent where the coefficients $a$ and $b$ (listed in
  Tab.~\ref{coeffs}) have been calculated -- by Herczeg \& Hillenbrand
  (\cite{herczeg}) -- by comparing optical emission line fluxes and
  accretion luminosities measured from the UV continuum excess. Errors
  on the coefficients $a$ and $b$ depend on the scatter of the line
  luminosity values around the best fit relation (from a large sample
  of sources), and therefore take into account several effects, such as
  optical depth and any contribution to the line emission from stellar
  outflows.

  Once the accretion luminosity, $L_{\mathrm acc}$, has been
  calculated, it is possible to calculate the mass accretion rate
  via:

\begin{equation}
\label{massacc}
\dot{M} = \left(1-\frac{R_{\mathrm{*}}}{R_{\mathrm{in}}}\right)^{-1} L_{\mathrm{acc}}\frac{R_{\mathrm{*}}}{G M_{\mathrm{*}}} \approx  1.25 L_{\mathrm{acc}} \frac{R_{\mathrm{*}}}{G M_{\mathrm{*}}}
\end{equation}

\noindent where $(1-R_{\mathrm{*}}/R_{\mathrm{in}})^{-1} \approx 1.25$
is estimated by assuming the accreting gas falls onto the star from
the truncation radius of the disk, $R_{\mathrm{in}} \approx
5R_{\mathrm{*}}$ (Gullbring et al. \cite{gullbring}).

We also calculated the mass accretion rate based on the H$\alpha$ full
width at 10\%, using the following equation (Natta et
al. \cite{natta04}):

\begin{equation}
\label{eq:halpha10}
\log \dot{M} \approx -12.9(\pm 0.3) + 9.7(\pm 0.7) \times 10^{-3}\, \mathrm{H}\alpha10\%  
\end{equation}

\noindent where $\mathrm{H}\alpha10\%$ is the H$\alpha$ 10\% full width in
km\,s$^{-1}$ and $\dot{M}$ is in M$_{\odot}$\,yr$^{-1}$. The derived
mass accretion rates, along with a (weighted) mean optical accretion
rate, $\langle\dot{M}_{\mathrm{Opt}}\rangle$, are listed in
Table~\ref{tab:massacc}. The mean optical accretion rate was
calculated using all the mass accretion rates calculated via
Eqs.~\ref{line}\, and~\ref{massacc}\, and therefore does not
include the mass accretion rate derived from the H$\alpha$ 10\% width
(see section~\ref{optcompsect} for more details).

\begin{table}
\caption{Coefficients used in Equation~\ref{line}, for each emission line, from Herczeg \& Hillenbrand (\cite{herczeg})}
\label{coeffs}
\centering
\begin{tabular}{lccc}
\hline \hline
Line & Wavelength & $a$ & $b$ \\
 & (\AA) & & \\
\hline
H$\gamma$ & 4349 & 3.0 $\pm$0.2 & 1.24 $\pm$ 0.04 \\
\ion{He}{ii} & 4686 & 3.7 $\pm$ 0.5 & 1.04 $\pm$ 0.08 \\
H$\beta$ & 4861 & 2.6 $\pm$0.2 & 1.22 $\pm$ 0.05 \\
\ion{He}{i} & 5016 & 3.3 $\pm$ 0.3 & 1.04 $\pm$ 0.05 \\
\ion{He}{i} & 5876 & 5.3 $\pm$ 0.7 & 1.46 $\pm$ 0.12 \\
\ion{O}{i} & 6300 & 2.8 $\pm$ 0.8 & 0.96 $\pm$ 0.16 \\
H$\alpha$ & 6563 & 2.0 $\pm$ 0.4 & 1.2 $\pm$ 0.11 \\
\ion{He}{i} & 6678 & 5.2 $\pm$ 0.8 & 1.37 $\pm$ 0.13 \\
\hline
\end{tabular}
\end{table}

\begin{sidewaystable*}
\vspace{15cm}
\caption{Derived optical mass accretion rates based on the equivalent widths and H$\alpha$ full width at 10\% measurements listed in Table~\ref{tab:linewidths}. Also listed are the optical weighted mean accretion rates, $\langle\dot{M}_{\rm Opt}\rangle$, based on all the accretion rates except H$\alpha$10\% width.}
\label{tab:massacc}
\centering
\begin{tabular}{lcccccccccc}
\hline \hline
\\
Source    & $\log \dot{M}_{\mathrm{H\gamma\ 4340\mbox{\AA}}}$ & $\log \dot{M}_{\ion{He}{ii}\ 4686\mbox{\AA}}$ & $\log \dot{M}_{\mathrm{H\beta\ 4861\mbox{\AA}}}$ & $\log \dot{M}_{\ion{He}{i}\ 5016\mbox{\AA}}$ & $\log \dot{M}_{\ion{He}{i}\ 5876\mbox{\AA}}$ & $\log \dot{M}_{\mathrm{H\alpha\ 6563\mbox{\AA}}}$ & $\log \dot{M}_{\mathrm{H\alpha\ 10\%}}$ & $\log \dot{M}_{\ion{O}{i}\ 6300\mbox{\AA}}$ & $\log \dot{M}_{\ion{He}{i}\ 6678\mbox{\AA}}$ &$\log \langle\dot{M}_{\mathrm{Opt}}\rangle$ \\
          & $M_{\odot}\,yr^{-1}$              & $M_{\odot}\,yr^{-1}$                   & $M_{\odot}\,yr^{-1}$             & $M_{\odot}\,yr^{-1}$                  & $M_{\odot}\,yr^{-1}$                  & $M_{\odot}\,yr^{-1}$                & $M_{\odot}\,yr^{-1}$           & $M_{\odot}\,yr^{-1}$                 & $M_{\odot}\,yr^{-1}$                 & $M_{\odot}\,yr^{-1}$ \\
\hline
Hen 3-600    & $-9.75\pm0.28$                     & $...$                                     & $ -9.61\pm0.30$                   & $ -9.52\pm0.42$                        & $ -9.02\pm0.94$                        & $ -8.89\pm0.55$                      & $ -10.15\pm0.61$                & $ -9.27\pm1.19$                       & $ -8.89\pm1.06$                      &  $-9.53\pm0.17$\\
TW Hya       & $-9.31\pm0.26$                     & $ -9.17\pm0.66$                         & $ -9.05\pm0.27$                   & $ -9.82\pm0.41$                        & $ -8.38\pm0.88$                        & $ -8.46\pm0.50$                      & $ -8.94\pm0.78$                 & $ -9.39\pm1.15$                       & $ -8.91\pm1.03$                      &  $-9.17\pm0.15$\\
RU Lup       & ...                                & ...                                     & $ -8.06\pm0.25$                   & $ -7.38\pm0.35$                        & $ -6.91\pm0.82$                        & $ -7.79\pm0.47$                      & $ -7.28\pm1.06$                 & $ -8.20\pm1.03$                       & $ -7.16\pm0.94$                      &  $-7.77\pm0.18$\\
BP Tau       & ...                                & $ -8.14\pm0.62$                         & $ -8.26\pm0.25$                   & $ -8.97\pm0.39$                        & $ -7.86\pm0.86$                        & $ -7.89\pm0.48$                      & $ -8.88\pm0.79$                 & $ -8.68\pm1.09$                       & $ -8.26\pm1.01$                      &  $-8.34\pm0.17$\\
V4046 Sgr    & ...                                & ...                                     & $ -9.34\pm0.28$                   & ...                                    & $ -9.07\pm0.91$                        & $ -8.79\pm0.51$                      & $ -7.89\pm0.95$                 & $ -9.62\pm1.17$                       & ...                                  &  $-9.22\pm0.23$\\
MP Mus       & $-9.61\pm0.26$                     & ...                                     & $ -9.32\pm0.27$                   & ...                                    & $ -8.89\pm0.90$                        & $ -8.73\pm0.50$                      & $ -8.01\pm0.92$                 & $ -9.18\pm1.11$                       & ...                                  &  $-9.35\pm0.17$\\
V2129 Oph    & ...                                & ...                                     & $ -9.76\pm0.29$                   & ...                                    & $ -9.19\pm0.92$                        & $ -9.28\pm0.55$                      & $ -10.23\pm0.60$                & $ -9.56\pm1.18$                       & $ -9.99\pm1.11$                      &  $-9.64\pm0.24$\\
T Tau        & ...                                & $ -7.90\pm0.60$                         & $ -7.22\pm0.23$                   & $ -7.50\pm0.35$                        & $ -6.98\pm0.81$                        & $ -7.16\pm0.44$                      & $ -8.82\pm0.80$                 & $ -7.68\pm0.97$                       & ...                                  &  $-7.32\pm0.16$\\
\hline
\end{tabular}
\end{sidewaystable*}

\subsection{X-ray Data Analysis}

\begin{figure*}
   \centering
   \includegraphics[width=16cm]{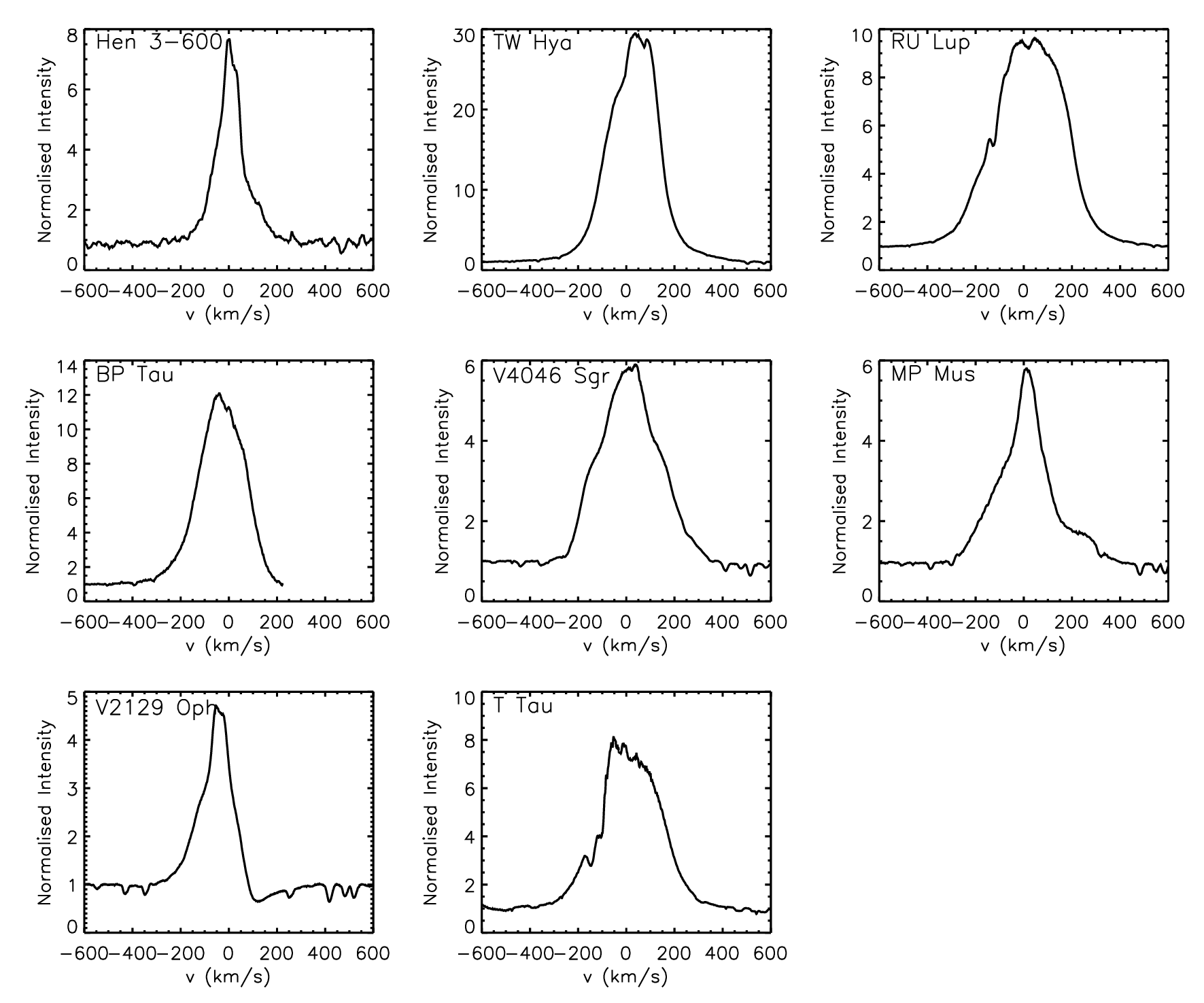}
      \caption{The observed H$\alpha$ line profiles for our sample. The
      data show many different line morphologies, with some very asymmetric
      lines e.g. V2129 Oph.}
         \label{halphalines}
   \end{figure*}

  

It is also possible to infer the mass accretion rate $\dot{M}$ of CTTSs
from their soft X-ray emission (e.g. Schmitt et al. \cite{schmitt}). This
method for deriving $\dot{M}$\, is based on the assumption that the
whole soft X-ray emission is due to accreting material and not to coronal
plasma, and that the post-shock zone can be described as a stationary
isothermal slab of plasma at constant velocity and density. Recent
hydrodynamical simulations (Sacco et al. \cite{germano}; \cite{sacco10}) have
demonstrated that, even if the post-shock zone is not stationary (and
characterized by quasi-periodic shock oscillations), the time-averaged
properties of the post-shock zone are, in general, well-described by
the stationary model. Sacco et al. (\cite{sacco10}) have also shown
that the higher the post-shock temperature the larger the discrepancies
between the stationary and the hydrodynamical models. This is mainly
due to the fact that the thermal conduction (which is not taken into
account in the stationary model and that is more efficient for higher
post-shock temperatures) drains energy from the shock-heated plasma to
the chromosphere through a thin transition region, and acts as an additional
cooling mechanism (see also Orlando et al. \cite{Orlando2010A&A}).

Letting $n_{\mathrm{0}}$ and $\vel_{\mathrm{0}}$ represent the
density and velocity of the pre-shock material, and $n_{\mathrm{1}}$
and $\vel_{\mathrm{1}}$ the corresponding post-shock values, then pre-
and post-shock quantities are linked by these relations:

\begin{equation}
\label{eq:shocks}
n_{\mathrm{1}} = 4 n_{\mathrm{0}},\ \
\vel_{\mathrm{1}} = \frac{1}{4}\vel_{\mathrm{0}},\ \
T_{\mathrm{1}} = \frac{3}{16}\frac{\mu m_{\mathrm{H}}}{k}
\vel_{\mathrm{0}}^{2}
\end{equation}

\noindent
where $T_{\mathrm{1}}$ is the post-shock temperature, $\mu$ is the
average mass per free particle ($\mu \approx 0.5$), and we assume a
strong shock regime. The volume occupied by the hot post-shock plasma
is defined by the stream cross-section, $A$, and by the distance,
$l$, covered by the plasma itself before cooling down. Hence, when
$\tau$ is the cooling time, $l$ is given by $\vel_{\mathrm{1}}\tau$,
and the emission measure -- $EM$ -- of the shock-heated plasma is
$n_{\mathrm{1}}^{2}A\vel_{\mathrm{1}}\tau$. Therefore it is possible to
derive the stream cross-section $A$, and hence the mass accretion rate,
from the $EM$ value measured from the soft X-ray emission via:

\begin{equation}
\dot{M} = \mu m_{\mathrm{H}} n_{\mathrm{1}} \vel_{\mathrm{1}} A 
        = \mu m_{\mathrm{H}} \frac{EM}{n_{\mathrm{1}}\tau} 
        = \mu m_{\mathrm{H}} \frac{EM\,P(T_{\mathrm{1}})}{3kT_{\mathrm{1}}}
\end{equation}

\noindent
where $\tau=3kT/(n_{\mathrm{1}}P(T))$, with $P(T)$ indicating the
plasma radiative losses per $EM$ unit. We used the above formula to
obtain $\dot{M}$ from the X-ray data. In particular for each CTTS: 1)
we evaluate $\vel_{\mathrm{0}}$, and hence $\vel_{\mathrm{1}}$, assuming
a free fall velocity from a distance of $5R_{\star}$ (to correspond
with the optical analysis); 2) we evaluate the post shock temperature
$T_{\mathrm{1}}$ from $\vel_{\mathrm{0}}$ (see Eq. \ref{eq:shocks});
3) the knowledge of $T_{\mathrm{1}}$ allows us to derive the plasma
$EM$ from the observed flux of the \ion{O}{vii} resonance line located
at 21.60\,\AA\ via:

\begin{equation}
EM=\frac{L_{OVII}}{G(T_1)\,A_{O}}
\end{equation}

\noindent
where $G(T_1)$ is the emissivity function of the \ion{O}{vii} resonance
line at 21.60\,\AA, and $A_{O}$ is the oxygen abundance.

We chose to rely our $\dot{M}$ estimations on the flux of the
\ion{O}{vii} resonance line because it is mostly produced by plasma at
2\,MK, which is the typical expected temperature for the plasma heated
in the accretion shock, and because its flux measurement is available
for almost all CTTSs observed with high resolution X-ray
spectroscopy.

Note that this method allows us to derive $\dot{M}$ independently of
the knowledge of the plasma density $n_{\mathrm{1}}$. On the other
hand this method depends on the plasma metallicity, which is needed to
calculate the radiative losses $P(T_{\mathrm{1}})$ and to link the
\ion{O}{vii} line flux to the corresponding $EM$. It is worth
  noting that X-ray observations provide us calibrated spectra, making
  the line flux measurements a simple and solid procedure, especially
  for the \ion{O}{vii} resonance line at 21.6\,\AA~that is an isolated
  line, and hence its flux estimation is not affected by problems due
  to line blending or wrong continuum placement. We decided,
  therefore, to infer the $\dot{M}$ values, with the above method,
  starting from published values of the fluxes of the \ion{O}{vii}
  resonance line.

In Table~\ref{tab:mdotx} we list, for each CTTS, the derived accretion
rates together with the main parameters used.

\begin{table*}
\caption{Observed X-ray parameters and derived X-ray mass accretion rates\tablefootmark{a}.}
\label{tab:mdotx}
\begin{center}
\begin{tabular}{lcccccccc}
\hline\hline

Name & $T$  & O, Ne, Fe & $P(T)$                                & $\log L_{\ion{O}{vii}}$       & $\log EM$         & $\log \dot{M}$ &  References for $L_{OVII}$\\
     & (MK) &           & $({\rm 10^{-23}cm^{3}\,erg\,s^{-1}})$ & $({\rm erg\,s^{-1}})$ & $({\rm cm^{-3}})$ & $({\rm M_{\odot}\,yr{-1}})$ & and abundances\\
\hline

Hen 3-600   & $1.0$ & $0.4,1.2,0.2$ & $5.7$ & $27.5 \pm 0.2$ & $ 52.7 \pm 0.2$ & $ -9.53 \pm 0.26$ & 1\\
TW Hya      & $3.1$ & $0.2,1.8,0.3$ & $2.2$ & $28.9 \pm 0.1$ & $ 53.4 \pm 0.1$ & $ -9.74 \pm 0.05$ & 2 \\
RU Lup      & $2.1$ & $0.6,1.4,1.1$ & $7.2$ & $28.9 \pm 0.1$ & $ 52.8 \pm 0.1$ & $ -9.67 \pm 0.13$ & 3 \\
BP Tau      & $1.8$ & $0.6,1.5,0.3$ & $4.7$ & $28.9 \pm 0.1$ & $ 52.8 \pm 0.1$ & $ -9.78 \pm 0.08$ & 4, 5 \\
V4046 Sgr   & $3.3$ & $0.1,1.0,0.1$ & $1.6$ & $28.5 \pm 0.1$ & $ 53.5 \pm 0.1$ & $ -9.84 \pm 0.11$ & 6, 7\\
MP Mus      & $4.1$ & $0.1,0.4,0.1$ & $1.3$ & $28.6 \pm 0.1$ & $ 53.8 \pm 0.1$ & $ -9.70 \pm 0.06$ & 8 \\
V2129 Oph   & ...   & ...           & ...   & ...            & ...             & ...         & ...\\
T Tau       & $2.9$ & $0.4,0.8,0.3$ & $2.4$ & $29.4 \pm 0.1$ & $ 53.6 \pm 0.1$ & $ -9.47 \pm 0.11$ & 9, 10\\

\hline
\end{tabular}
\end{center}
\tablefoottext{a}{Chandra data for V2129 Oph have recently been obtained
(Argiroffi et al. in prep) but not published yet.}
\tablebib{(1) Huenemoerder et al. \cite{huenemoerder}; (2) Stelzer \& Schmitt, \cite{beate}; (3) Robrade \& Schmitt \cite{robrade}; (4) Schmitt et al. \cite{schmitt}; (5) Robrade \& Schmitt \cite{robrade06}; (6) G\"unther et al. \cite{gunther}; (7) Argiroffi C., private communication; (8) Argiroffi et al. \cite{costanza09}; (9) G\"udel \& Telleschi \cite{gudeltell}; (10) G\"udel et al. \cite{gudel}.}
\end{table*}

\section{Results \& Discussion}
\subsection{Comparison of optical derived accretion rates}
\label{optcompsect}

Table \ref{tab:massacc} shows the calculated mass accretion rates for
each star based on the different emission lines analysed. Although the
sources were observed at different epochs, all the emission lines were
observed simultaneously for each star. Therefore one might na\"{i}vely
expect the different tracers to lead to very similar mass accretion
rates (if each line is tracing the same accretion phenomena). The top
panel of Fig.~\ref{opticalcomp} shows the different mass accretion
rates calculated for each star in the sample (in order of
  increasing stellar mass). We find that all the different tracers
agree within the errors, albeit with a spread of typically $\sim$1
order of magnitude. We see no relation between mass accretion
  rate and the mass of the stars. Also, it should be noted, that there
  is no apparent relation between the mass accretion rate and the age
  of the stars (see table~\ref{param}) -- for example, V2129 Oph has
  an age of 2 Myrs, whereas T Tau has an age of 2.7 Myrs and their
  mass accretion rates are vastly different. V4046 Sgr has an age of
  12 Myrs and MP Mus has an age of 6--7 Myrs, and yet they have
  similar mass accretion rates. The bottom panel of
Fig.~\ref{opticalcomp} shows the distribution of the difference
between each mass accretion rate estimate and
$\langle\dot{M}_{\mathrm{Opt}}\rangle$, the optical mean mass
accretion rate. In general, H$\alpha$ flux and \ion{He}{i} 5876\AA\,
lead to higher mass accretion rates than H$\beta$, H$\gamma$ and
\ion{O}{i}.

H$\alpha$ 10\% mass accretion rates are often at the extremes of the
range of estimates, but this is not surprising given the H$\alpha$
line profiles are dependent on inclination, rotation effects (see
Mohanty et al. \cite{subu05} for details), winds, and
outflows. Fig.~\ref{halphalines} shows the H$\alpha$ line profiles for
our sample, some of which are very asymmetric. The H$\alpha$ emission
from Hen 3-600 and MP Mus both have narrow peaks, but with asymmetric
broad wing emission. TW Hya, RU Lup and BP Tau all have broader peaked
emission, albeit with some evidence for self-absorption. V4046 Sgr has
an H$\alpha$ line profile which has a broad peak, with very broad wing
emission. V2129 Oph and T Tau both show clear evidence for absorption
in the H$\alpha$ emission, with V2129 Oph having a very asymmetric
profile with strong absorption of the blue-shifted wing/no blueshifted
component. Interestingly, we see that the accretion rates derived from
H$\alpha$ 10\% emission is to the lower part of the range of mass
accretion rates for Hen 3-600, BP Tau, V2129 Oph and T Tau, at the
upper end of the range for RU Lup, V4046 Sgr and MP Mus and in the
middle of the range for TW Hya (see top panel of
Fig.~\ref{opticalcomp}). The asymmetry of the emission line profile
from V2129 Oph would clearly lead to the under-estimation of the mass
accretion rate for that object, but interestingly, the
under-estimation is not extreme -- the difference between the
H$\alpha$ 10\% mass accretion rate and the optical mean mass accretion
rate being only $-1.7 \times 10^{-10}$ M$_{\odot}$\,yr$^{-1}$. In the
cases of V4046 Sgr, MP Mus and T Tau, these estimates are clear
outliers from the rest of the accretion rate estimates (high estimates
for the first two -- as predicted from the line profiles, and lower
for T Tau -- which has a `normal' line profile, albeit with some
absorption, but this does not affect the 10\% intensity level).  Due
to the fact that H$\alpha$ 10\% mass accretion rate estimates are
outliers in almost half our sample, along with the fact that
inclination, rotation, winds and outflows may affect these mass
accretion rates, we have excluded the H$\alpha$ 10\% mass accretion
rate estimates from the calculation of the optical mean mass accretion
rate listed in Table~\ref{tab:massacc}.

\subsection{Variability of TW Hya} 

\begin{figure}
\centering
\includegraphics[width=8cm]{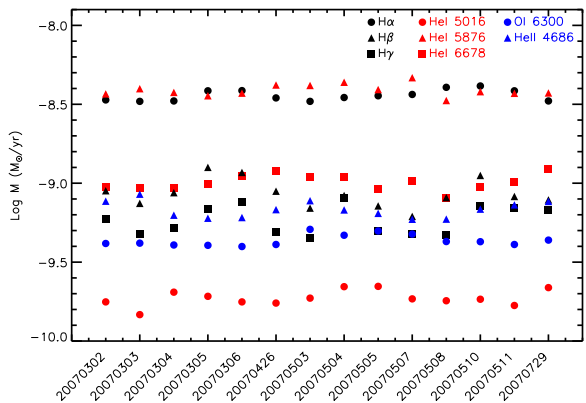}
\caption{The variability of TW Hya. The different
  color/symbols represent different accretion rate tracers (see
  upper right corner). Note the X-axis timescale is not linear. The
  variability in accretion rate is observed to be small in comparison
  to the range of accretion rates calculated from the different tracers.}
\label{variability}
\end{figure}

TW Hya is the only source within the sample to have been observed over
a period of months (from March 2007 through to July 2007) on both
short (nightly) and long (monthly) timescales, using the same
instrumental set-up. We have analysed 14 nights of data over a 5 month
period for the purpose of studying the accretion variability. Our
results are shown in Fig.~\ref{variability}. We find the variability
for a given tracer small compared to the $\sim$ 1 order of magnitude
spread in mass accretion rates from the different tracers. \ion{O}{i},
H$\alpha$ and \ion{He}{i} 5876\AA\, are the tracers with the least
variability, whilst H$\beta$ and H$\gamma$ have the largest
variability. It could be argued that this small variability is
expected given TW Hya is viewed almost pole-on, and as such it should
be possible to observe all active accretion regions about the facing
pole at all times, therefore eliminating large changes in mass
accretion rates as the hotspots rotate out of view. The variability of
mass accretion rates are not necessarily due only to the inclination
angle of the star-disk system, but may be due to time-dependent
accretion rates. The small amount of variability ($\sim$ 0.25 order of
magnitude) we observe in our TW Hya data may be due to time-dependent
accretion rate.

Alencar \& Batalha (\cite{alencar}) studied the TW Hya accretion rate
for 1.5 yrs, and found the mass accretion rate to be variable between
$10^{-9}$ and $10^{-8}$ M$_{\sun}$\,yr$^{-1}$\, using the \ion{Na}{d}
line profile. Whilst these results fall within the range of mass
accretion rates we measured using different tracers, they suggest a
larger variability. This discrepancy may be due to the method used to
estimate the mass accretion rate or due to the longer timeperiod
covered in their analysis. Rucinsky et al. (\cite{rucinsky}) analysed
two datasets -- from 2007 and 2008 -- and found a photometric
variability of $\Delta V < 0.5$ mag on timescales of $\sim$ 3.5 days
in the 2007 data. This however was no longer observable in the 2008
data, which showed variability (again $\Delta V < 0.5$ mag) on
timescales of 2--9 days, which they suggest may be due to the orbital
decay of accreting gas. In addition, they find `spikes' in the
accretion rate lasting only a fraction of a day. They conclude that
whilst accretion contributes to this variability, it is not the sole
cause.

  Our optical and X-ray observations are not coeval, and so it is
  important to study the variability of these stars due to accretion
  to know if the their variability will limit our ability to compare
  mass accretion rates derived from non-coeval observations. We find
  TW Hya, a pole-on star, to have a variability within our
  errors. Furthermore, all the sources in this sample have
  inclinations $\lesssim$ 45$\degr$, with TW Hya, Hen 3-600, RU Lup
  and T Tau all having inclinations much closer to pole-on, meaning
  any accretion hotspots would be visible all/most of the time,
  limiting variability due to observability of the hotspot.Whilst we
  cannot rule out time dependent mass accretion rates for the stars
  other than TW Hya in our sample, we measure a small variability in
  our TW Hya data, and even the larger variabilities suggested in the
  literature fall within our errors.

\subsection{Comparison of optical/x-ray derived accretion rates}

 \begin{figure*}
   \centering
   \includegraphics[width=16cm]{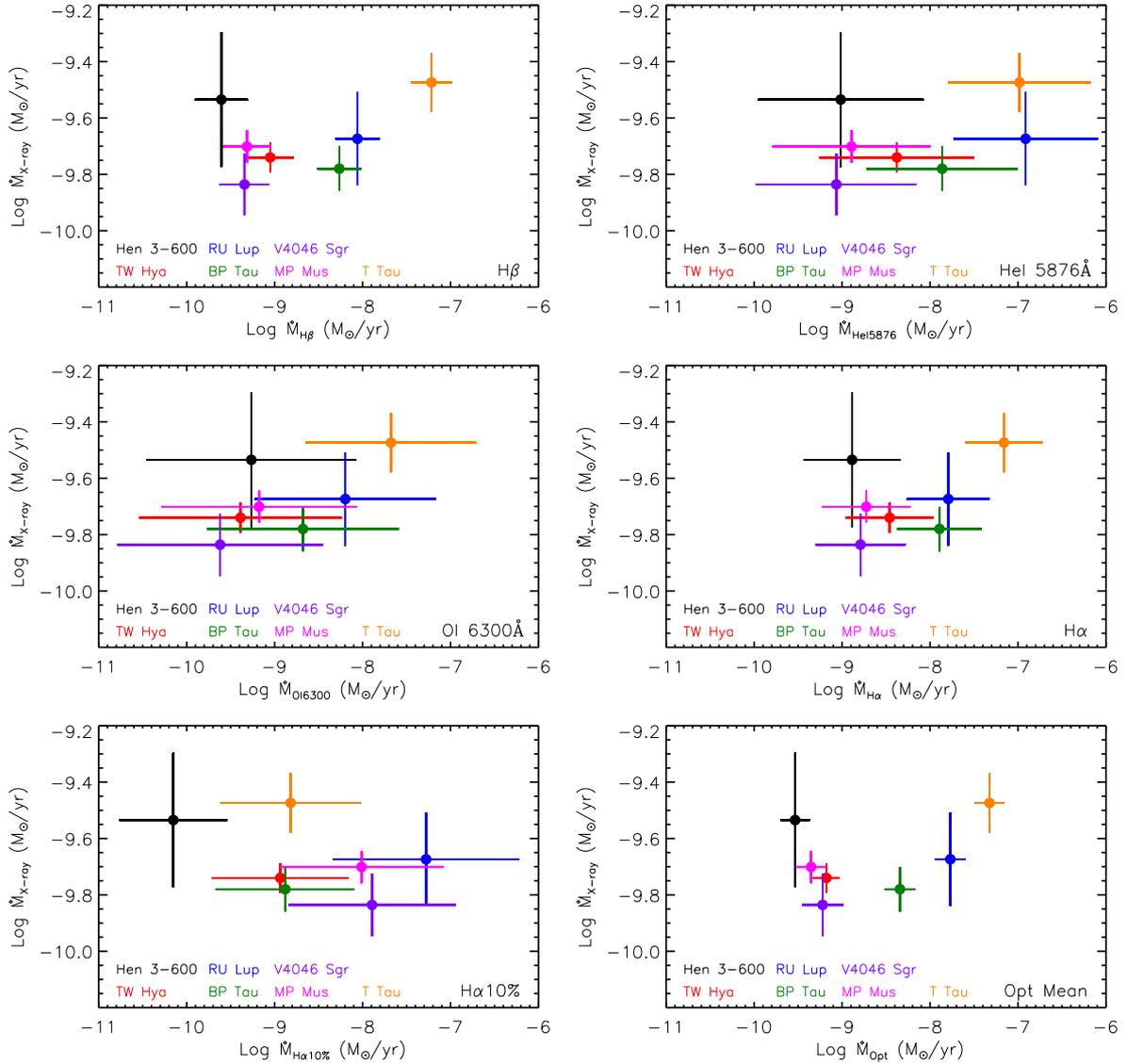}
   \caption{Plots of the optically derived mass accretion rates versus
     the X-ray derived mass accretion rates. Each panel shows the
     optical accretion rate derived from a different tracer, indicated
     in the lower right corner of the plot. The CTTSs are colour coded
     (see lower left corner of each panel). We plot only those
     cases for which we have measured the mass accretion rate for all
     the stars in our sample (see table~\ref{tab:massacc}). The bottom
     right panel shows a plot of the optical mean mass accretion rate
     plotted against the X-ray accretion rate. There is a very small
     range, a factor $\approx 2$, in X-ray accretion rate
     calculated for the entire sample, whereas the optical mass
     accretion rates span a range of $\sim 3$ orders of magnitude.}
         \label{opt-xray}
   \end{figure*}

We compare the optical derived mass accretion rates $\langle\dot{M}_{\rm
Opt}\rangle$ to the X-ray derived mass accretion rates $\dot{M}_{\rm
X-ray}$ in Fig.~\ref{opt-xray}. The most striking result is
that $\dot{M}_{\rm X-ray}$ for our sample ranges within a factor
$\approx 2$ around $2 \times 10^{-10}$ M$_{\odot}$\,yr$^{-1}$ (see
Fig.~\ref{opt-xray} and Table~\ref{tab:mdotx}), despite the fact that
the range of $\langle\dot{M}_{\rm Opt}\rangle$ spans almost 3 orders of
magnitude. In addition, as suggested by other studies in the literature,
the X-ray mass accretion rates are always lower than the corresponding
optical mass accretion rates with the discrepancy increasing for
increasing $\langle\dot{M}_{\rm Opt}\rangle$.

The above results can be interpreted in the light of the findings of Sacco
et al. (\cite{sacco10}) who used a detailed hydrodynamic model
to investigate the observability of accretion shocks in X-rays in a wide
range of physical conditions of the accretion stream. They considered the
absorption from the optically thick plasma of the stellar chromosphere
on the X-ray emission arising from the shock-heated material. They
found that the absorption strongly depends on the accretion stream
properties and influences the observability of the post-shock accreting
plasma in the X-ray band. In particular, observable X-ray emission from
accretion shocks is expected in streams with densities in the range
$10^{11}\lesssim n_{\rm e}\lesssim 10^{12}$ cm$^{-3}$ and velocities in
the range $400\lesssim \vel\lesssim 600$ km\,s$^{-1}$. In fact,
denser streams produce post-shock zones that are rather narrow and deeply
rooted in the chromosphere, causing their X-ray emission to be strongly
absorbed. Streams with velocities below 400 km\,s$^{-1}$ produce shocked
plasma with temperatures around 1 MK or even lower and, therefore, hardly
observable in X-rays. The obvious consequence is that, whilst accretion
shocks are almost always observable in the optical band, only a small
fraction of them can be revealed in X-rays, leading in general to an
underestimate of the X-ray derived mass accretion rate and, therefore,
to the observed discrepancy with the optical derived mass accretion
rates (see also discussion in Orlando et al. \cite{Orlando2010A&A}
and Sacco et al. \cite{sacco10}).

Note that the above argument is applicable even if only one stream
is present on the star. In fact, for typical densities and velocities
of accretion streams and stellar magnetic field strengths (as derived
from observations of CTTSs; Johns-Krull \cite{johnskrull}), in general
the plasma $\beta$ (i.e. the ratio gas pressure / magnetic pressure)
is expected to be $\ll 1$ (see Sacco et al. \cite{sacco10}). In these
conditions, an accretion stream is a bundle of fibrils each independent
of the others due to the strong magnetic field which prevents mass
and energy exchange across magnetic field lines (e.g. Orlando et
al. \cite{Orlando2010A&A} and Sacco et al. \cite{sacco10}). If the stream
is structured in density (as suggested, for instance, by 3D MHD models of
the star-disk system carried out by Romanova et al. \cite{romanova04}),
only fibrils with density and velocity in the range of observability
defined by Sacco et al. (\cite{sacco10}) are expected to produce
detectable X-rays, unless they are located close to the centre of
the stream cross-section where they suffer strong absorption from the
accretion column itself.

Interestingly, Hen 3-600 is the lowest mass star in our sample and has
a close agreement between the optical and X-ray mass accretion rates,
suggesting it has very few high density streams (or fibrils) --
hardly detectable in X-rays -- or only low density accretion streams
(fibrils) -- observable in X-rays -- meaning almost everything
is seen in the X-ray, as well as in the optical. At the other end of
the scale, RU Lup could be interpreted to have a much
larger number of high density streams (fibrils).

If, on one hand, the discrepancy between optical and X-ray derived mass
accretion rates may be easily explained in the framework of accretion
shock models as discussed above, on the other hand, more puzzling is the
result that $\dot{M}_{\rm X-ray}$ is almost the same for all the stars
of our sample. This result is even more surprising considering that
the determination of $\dot{M}_{\rm X-ray}$ assumes some fixed parameters
(e.g. the distance, the mass and the radius of the star, the abundance
of the emitting plasma, and the temperature of the post-shock region)
whose uncertainties are unknown and therefore not included in the errors
on $\dot{M}_{\rm X-ray}$. Taking into account all these uncertainties,
one would expect a scatter in the values of $\dot{M}_{\rm X-ray}$
possibly larger than one order of magnitude. On the contrary, the
evidence is that the observed scatter is within a factor $\approx 2$
and it can be explained only if the (unknown) uncertainties on the fixed
parameters used in our calculations are smaller than expected.

As for the almost constant value of $\dot{M}_{\rm X-ray}$, we note
that its estimate relies on the flux of the \ion{O}{vii} resonance line
and is based on the assumption of optically thin plasma. However it can
be inferred that optical depth of the strongest emission lines produced
by the plasma located in the post-shock region is non-negligible. In fact
for a post shock region with dimension $\sim$10$^9$ cm, filled with plasma
at T=2 MK and n$_{\rm e}$=10$^{11}$ cm$^{-3}$, the optical depth of the
\ion{O}{vii} resonance line is $\sim 10$. Supporting this scenario is some
evidence of optical depth effects (i.e. the emitting plasma is not totally
optically thin) which has been detected in the X-ray spectrum of MP Mus
(Argiroffi et al. \cite{costanza09}), implying that the \ion{O}{vii}
flux and the derived $\dot{M}_{\rm X-ray}$ could be underestimated.

The optical depth effects can be addressed by comparing mass
accretion rates measured from the resonance line of the \ion{O}{vii}
triplet with the accretion rate measured using the flux of the
intercombination plus the flux of the forbidden line of the same triplet,
which are less affected by optical depth effects. We performed this
comparison in the case of TW Hya and MP Mus. We found no significant
differences in the former case, while the $\dot{M}_{X-ray}$ of MP Mus
measured from the intercombination plus the forbidden line is twice as
large as the $\dot{M}_{X-ray}$ measured from the resonance line. The lack
of high signal-to-noise spectra does not allow us to further investigate
this issue using this approach for the other stars in our sample. However,
this preliminary result suggests that the absorption from the stellar
atmosphere is the main cause of the discrepancy between the optical and
the X-ray mass accretion rates.

Brickhouse et al. (\cite{brickhouse}) analyzed X-ray spectroscopic
data of TW Hya with high signal-to-noise ratio.  This kind of analysis
is limited to the case of TW Hya, which is the brightest CTTS in the
X-ray and has been observed with Chandra for $\sim$500 ks. These
authors found that the flux and the densities measured from the
\ion{Ne}{ix} triplet agree with the accretion shock parameters derived
from optical emission lines, while the density and flux derived from
the \ion{O}{vii} triplet disagree.  They proposed therefore a model of
``accretion-fed corona'' in which the X-ray emission originates from
three plasma components: a hot ($T_{\rm e} \approx 10$ MK) corona, a
high density ($n_{\rm e} \approx 6.0 \times 10^{12}$ cm$^{-3}$,
$T_{\rm e} \approx 3.0$ MK) post-shock region close to the shock
front, and a cold less dense ($n_{\rm e} \approx 2 \times 10^{11}$
cm$^{-3}$, $T_{\rm e} \approx 2.0$ MK) post-shock cooling region, with
300 times more volume and 30 times more mass than that of the post
shock region itself. On the other hand, it is worth to note that,
recently, Sacco et al. (\cite{sacco10}) have shown that, if accretion
streams are inhomogeneous and the chromospheric absorption is
efficient, their hydrodynamic model of accretion shock predicts that
different He-like triplets measure different densities of the X-ray
emitting plasma. This is explained because the effects of absorption
increases with wavelength and, as a consequence, the \ion{Ne}{ix}
(13.45 {\AA}) emission results to be less absorbed than \ion{O}{vii}
(21.60 {\AA}) emission. In particular, Sacco et al. (\cite{sacco10})
showed that, in an inhomogeneous stream affected by chromospheric
absorption, the average density measured with the \ion{Ne}{ix} triplet
is, in general, larger than that measured with \ion{O}{vii} triplet
although \ion{Ne}{ix} lines form at temperatures higher than
\ion{O}{vii} lines (as noted by Brickhouse et
al. \cite{brickhouse}). Unfortunately, due to the lack of high
signal-to-noise data covering the \ion{Ne}{ix} triplet, we are not
able to test this model for the other stars of our sample. Further
long exposure X-ray spectroscopic observations of the stars in our
sample are required to test this model as well as optical depth
effects on the estimate of $\dot{M}_{X-ray}$.

If the results presented in this paper are confirmed, we suggest
here that the optical depth effects in the \ion{O}{vii} resonance line may
explain the almost constant $\dot{M}_{\rm X-ray}$ derived in our sample if
the effect is larger in stars with larger optical derived mass accretion
rates. Finally, it is worth to note that our analysis does not allow
us to conclude that there is no relation between $\langle\dot{M}_{\rm
Opt}\rangle$ and $\dot{M}_{\rm X-ray}$. From our analysis we can just
conclude that the relation, if exists, is masked by the (small) scatter in
the values of $\dot{M}_{\rm X-ray}$. To this respect, it is interesting
to note that the lower right panel in Fig.~\ref{opt-xray} seems to
suggest that $\dot{M}_{\rm X-ray}$ slightly increases for increasing
$\langle\dot{M}_{\rm Opt}\rangle$ if we consider that the value of
$\dot{M}_{\rm X-ray}$ for Hen 3-600 shows the largest error in X-rays.

\section{Conclusions}

We have carried out the first homogeneous comparison of optical and
X-ray derived mass accretion rates for a sample of CTTSs. We compare
the different optical tracers to each other, and to the X-ray derived
accretion rates. We also analysed the variability of the CTTS TW
Hya. Our findings lead to the following conclusions:

\begin{enumerate}

\item The mass accretion rates derived from the different optical
  tracers agree within the errors for each source, albeit with a large
  spread of typically $\approx 1$ order of magnitude (see
  Fig.~\ref{opticalcomp}).

\item H$\alpha$ 10\% full width, whilst known to be a good {\em
  indicator} of accretion, is not good at {\em measuring} mass
  accretion rates.

\item For the CTTS TW Hya (the only source within the sample for
  which a time analysis was possible), we find little variation in the
  mass accretion rates (within each emission line tracer) over a period
  of 5 months (see Fig.~\ref{variability}). The variability is
  much smaller than the range of accretion rates derived from different
  accretion tracers.

\item The X-ray mass accretion rates are always smaller than the
  optically derived mass accretion rates for all sources within
  the sample (see Fig.~\ref{opt-xray}). This can be explained if the
  accretion streams are inhomogeneous and/or multiple streams with
  different densities are present simultaneously. In these cases, Sacco
  et al. (\cite{sacco10}) have shown that the chromospheric absorption triggers
  a selection of the X-ray emitting shocks, absorbing preferentially
  the X-ray emission from high density plasma components. As a result,
  only the low density plasma component of the post-shock plasma can
  be observed in the X-ray band, leading to a systematic underestimate
  of the mass accretion rate.

\item We find that the X-ray derived mass accretion rate
  ranges within a factor of $\approx 2$ around $2 \times 10^{-10}$
  M$_{\odot}$\,yr$^{-1}$ (see Fig.~\ref{opt-xray}), despite the fact
  that the range of optical mass accretion rates span almost 3 orders of
  magnitude. Some evidence of non-negligible optical depth of emission
  lines produced by post-shock accreting plasma (e.g. Argiroffi et al.
  \cite{costanza09}) may explain the almost constant $\dot{M}_{\rm X-ray}$
  if the effect is larger in stars with larger optical mass accretion
  rates. This issue deserves further investigation in future studies to
  assess the severity of optical depth effects on the estimate of mass
  accretion rates in the X-ray band.

\end{enumerate}

\begin{acknowledgements}
  Based on observations made with ESO Telescopes at the La Silla or
  Paranal Observatories under programme ID $<069.C-0481>$, $<073.C-0355>$,
  $<075.C-0292>$, $<077.D-0478>$, $<078.A-9059>$, $<079.A-9006>$,
  $<079.A-9007>$ and $<079.A-9017>$. Based on observations made with the
  Italian Telescopio Nazionale Galileo (TNG) operated on the island of
  La Palma by the Fundaci\'{o}n Galileo Galilei of the INAF (Istituto
  Nazionale di Astrofisica) at the Spanish Observatorio del Roque de
  los Muchachos of the Instituto de Astrofisica de Canarias. This
  work was supported by the EU Marie Curie Transfer of Knowledge program
  PHOENIX under contract No. MTKD-CT-2005-029768 and in part by Agenzia
  Spaziale Italiana under contract No. ASI-INAF I/088/06/0.
\end{acknowledgements}

\end{document}